\documentclass{article}
\pdfoutput=1

\usepackage[preprint, nonatbib]{tackling_climate_workshop_style}

\usepackage[utf8]{inputenc} 
\usepackage[T1]{fontenc}    
\usepackage{float}
\usepackage[unicode, pdfusetitle]{hyperref}       
\usepackage{url}            
\usepackage{booktabs}       
\usepackage{microtype}      

\usepackage{amsmath}
\usepackage{amssymb}
\usepackage{mathtools}
\usepackage{nicefrac}       
\usepackage{siunitx}

\usepackage{caption}
\usepackage{subcaption}

\usepackage{graphicx}
\usepackage{todonotes}

\usepackage[
    style=numeric,
    backend=bibtex,
    url=false,
    sorting=none
]{biblatex}
\AtBeginBibliography{\small}

\bibliography{bibliography}

\DeclareMathOperator*{\argmin}{arg\,min}
\newcommand{\Lc}{\mathcal{L}}
\DeclareMathOperator*{\rmse}{rmse}
\newcommand{\ts}{\textsuperscript}

\title{A locally time-invariant metric for climate model ensemble predictions of extreme risk}

\author{%
    Mala Virdee \\
    University of Cambridge\\
    \texttt{mv490@cam.ac.uk} \\
    \And
    Markus Kaiser \\
    University of Cambridge, \\
    Monumo \\
    \texttt{mk2092@cam.ac.uk} \\
    \And
    Carl Henrik Ek \\
    University of Cambridge \\
    \texttt{che29@cam.ac.uk} \\
    \And
    Emily Shuckburgh \\
    University of Cambridge \\
    \texttt{efs20@cam.ac.uk} \\
    \And
    Ieva Kazlauskaite \\
    University of Cambridge \\
    \texttt{ik394@cam.ac.uk}
}

\begin{document}

\maketitle


\begin{abstract}Adaptation-relevant predictions of climate change are often derived by combining climate model simulations in a multi-model ensemble.
Model evaluation methods used in performance-based ensemble weighting schemes have limitations in the context of high-impact extreme events. We introduce a locally time-invariant method for evaluating climate model simulations with a focus on assessing the simulation of extremes.
We explore the behaviour of the proposed method in predicting extreme heat days in Nairobi and provide comparative results for eight additional cities.
\end{abstract}


\section[Introduction]{Introduction}
\subsection{Background}

Climate change is increasing the frequency and severity of extreme weather events, including high-temperature extremes (\cite{portner2022climate}).
The occurrence of heat extremes exceeding human heat stress thresholds is associated with increased mortality and morbidity, particularly in rapidly urbanising developing economies (\cite{tuholske2021global}).
People particularly exposed and vulnerable to heat stress risk include the urban poor, those in informal housing, the elderly, those with chronic health conditions, and outdoor workers (\cite{cardona2012determinants}).
Reliable predictions of future changes in the frequency, intensity and distribution of high-temperature extremes are particularly critical for cities --- impacts are amplified by urban heat island effects and high population density, but city-scale adaptation measures have been demonstrated to significantly reduce risk (\cite{estrada2017global}). 
Here, we predict extreme heat days, defined as days on which average temperature exceeds the 90\ts{th} percentile of local historically observed temperatures in accordance with several other analyses of changing extreme heat risk (\cite{seneviratne2014no, morak2013detectable}).
We introduce a method of evaluating the skill of climate models in simulating observed extreme heat days and derive a multi-model ensemble scheme with a focus on predicting these extremes. 

The latest General Circulation Models (GCMs) effectively reproduce observed large-scale trends and provide robust predictions of global average changes, but exhibit significant uncertainty in the local regime and for prediction of extremes (\cite{flato2014evaluation}).
A growing sector of `climate services' aims to bridge the gap between seasonal local weather forecasting and long-term mean climatology to provide decadal to multi-decadal predictions for use in impact assessment and development of adaptation strategies (\cite{meehl2009decadal}).
Deriving decision-relevant information from GCMs typically involves combining predictions from several models in a multi-model ensemble.
The multi-model approach aims to provide more skilful and robust predictions by utilising the various strengths of different models, as well as an estimate of structural model uncertainty (\cite{stainforth2007issues}). 

The most straightforward and widely-used method of combining predictions from multiple models is to calculate a multi-model mean, which has been found to outperform any individual model for a range of tasks (\cite{weigel2008can}).
However, an equally-weighted ensemble does not take into account model skill in simulating the historically observed quantity of interest, and assumes each model is an independent estimate.
Modelling groups often share assumptions and biases, leading to overconfident predictions (\cite{tebaldi2007use}).
Alternative methods involving unequal independence-based or skill-based weighting of ensemble members include Reliability Ensemble Averaging (\cite{giorgi2003probability}), Independence-Weighted Mean (\cite{bishop2013climate}), and Bayesian Model Averaging (BMA) (\cite{raftery2005using}). \par
In Section \ref{section:methodology} we introduce a novel method for the evaluation of model simulations which optimises a skill-based weighting.
Here, BMA is used to derive a predictive probability distribution of future temperatures by assigning skill-based weights to each climate model. 
The approach to model evaluation proposed here could be incorporated into any other skill-based model weighting approach --- for a review of schemes used to derive probabilistic predictions from climate model projections for impact assessment and adaptation planning, see \cite{brunner2020comparing}. 

\subsection{Related work}
The problem addressed in this work is the measurement of similarity between a simulated and observed time-series in the context of climate model evaluation. 
A similar problem is faced in many other contexts where comparing two time-series either according to Euclidean distance or by comparing summary statistics is found to be insufficient.
Dynamic Time Warping (DTW) algorithms allow non-linear alignment between series in contexts where an informative measure should consider the similarity between the `shape' of two signals rather than local temporal synchronisation. 
For example, in the field of automatic speech recognition where DTW originates, a measure should register similarity between the same speech pattern spoken at different speeds (\cite{rabiner1993fundamentals}). In climatology, DTW has been applied to measure similarity between local climates to develop a global climate classification scheme (\cite{netzel2017world}). 

As discussed further in Section \ref{section:methodology}, strictly order-preserving alignments such as DTW and its extensions lack flexibility. An alternative approach is to use a divergence metric to calculate a distance between two distributions --- disregarding the temporal structure of data allows extremes to be compared. Optimal Transport (OT) based divergence methods have been applied to climate model evaluation, for instance ranking models according to Wasserstein distance from observed climate (\cite{vissio2020evaluating}).

Some existing methods share the motivation of the work presented in this paper towards developing a flexible time-series similarity measure that also incorporates temporal structure (\cite{zhang2020time}). However, the authors are not aware of any application of such methods for climate model evaluation.

\section[Methodology]{Methodology} \label{section:methodology}
\subsection{Model evaluation}

\begin{figure}[h!]
    \centering
    \includegraphics{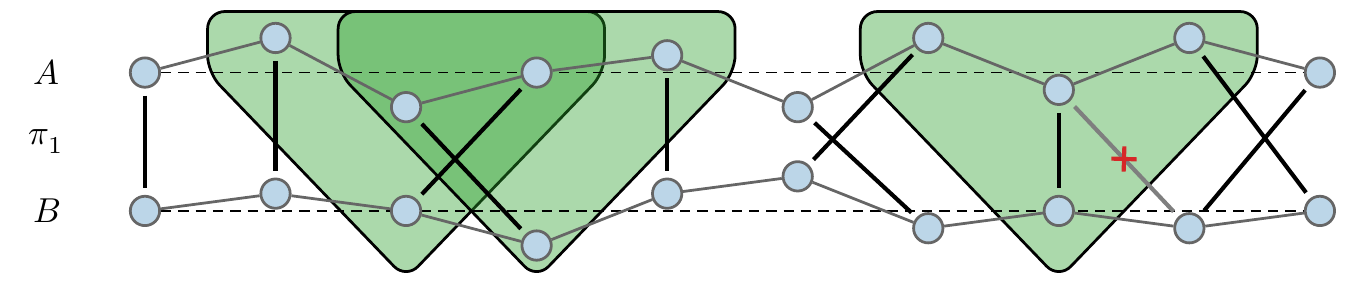}
    \caption{
        The locally time-invariant skill metric $\Lc$ is used to compare a simulated time-series $A$ to a reference time-series $B$.
        Instead of calculating the pairwise least-squares error, we propose adding a slack in either direction for reference points with which each simulated data-point can be matched.
        In this illustration a slack of one time-step in either direction is added, as represented by the green shapes.
        We then find an optimal bipartite matching $\pi$ that minimises the sum of distances between the time-series.
        On the left, data-points are compared out-of-order in overlapping windows to calculate distance between $A$ and $B$.
        On the right, we emphasise that the bipartite matching enforces the constraint that no data-point in either time-series can be used twice
        \label{fig:matching_metric}
        }
\end{figure}

Since GCMs simulate climate, they are not expected to provide synchronous simulations of weather at a specific location under future climate conditions, but it is assumed that they can yield informative statistics of future weather at some aggregate scale.
Point-wise evaluation of daily simulations against historical observations requires a climate model to predict weather. 
Given sequences of $T$ daily simulations from a climate model $A$ and historical observations $B$ against which they are to be evaluated, we cannot expect the time-series to match under a daily point-wise error measure such as the root mean squared error $\rmse(A, B) \coloneqq \sqrt{\frac{1}{T}\sum_{t=0}^{T-1} (A_t - B_t)^2}$. 

We expect this error to be high even for GCMs that are skilled in reproducing large-scale patterns. RMSE implicitly assumes the two time-series to be aligned by comparing the predictions for individual days $t$.
Summary statistics aim to avoid this issue by introducing some degree of time-invariance by binning data into monthly, seasonal or longer periods. Model error can then be calculated per time-bin, for instance by comparing the simulated and observed average or variance for each period, or by counting-based error measures such as comparing simulated and observed histograms to assess simulated variability.
Within each bin, simulations could be permuted freely in time to yield the same measure of skill.
This enables models to be evaluated without placing the expectation that they should simulate weather. 

However, these summary statistics share two problems.
First, they introduce artificial time boundaries by binning into periods to be evaluated independently.
It is unclear how boundaries should be chosen optimally (\emph{e.g.}, at the start or middle of each month) and in a way that minimises loss of accuracy introduced by rapidly changing weather conditions.
Second, model precision is reduced. This method of introducing time-invariance blurs the model outputs to the resolution of the bins.
This is problematic for localised extreme event prediction tasks, where retaining precision may be important. 

We propose an evaluation method to reduce the inaccuracy introduced through summary statistics whilst conserving the time-invariance required to avoid the implicit requirement to predict weather.
We assume a \emph{weather window size $w$} of time steps within which we cannot expect simulated data-points to be aligned with observed data-points.
To construct a metric that is \emph{locally time-invariant}, we introduce a permutation $\pi_w$ to the standard RMSE-measure before calculating differences in
\begin{align}
\Lc^w_{\pi_w}(A, B) & \coloneqq \sqrt{\frac{1}{T} \sum_{t=0}^{T-1} (A_t - B_{\pi_w(t)})^2}.
\end{align}
The permutation is constrained to locally reorder the time-series within the weather window - that is, every $A_t$ can be compared to the values between $B_{t-w}$ and $B_{t+w}$.
Note that this construction is symmetric with respect to $A$ and $B$.
The final metric $\Lc^w$ is given by choosing the locally constrained permutation that minimises the RMSE in
\begin{align}
    \begin{split}
        \Lc^w(A, B)             & \coloneqq \Lc^w_{\pi_w^\ast}(A, B)                                   \\
        \text{with } \pi_w^{\ast} & \in \argmin_{\pi_w} \Lc^w_{\pi_w}(A, B).
    \end{split}
    \label{eq:2}
\end{align}
Intuitively, we compare the simulations with observations under the assumption that the model was able to predict the weather as well as possible.
See Figure \ref{fig:matching_metric} for a graphical representation of the algorithm.
Since $\pi_w$ is a permutation, the data-points in either time-series can only be used once, preventing the metric from inventing new data. 

Introducing local time-invariance through $\pi_w$ solves a similar problem to calculating summary statistics by binning, but it is more precise:
by design, there are no boundaries of bins since the whole time-series can be considered at once.
As a consequence, the weather window size $w$ can be chosen to be smaller than a bin width since no effects at the boundaries or effects due to bad placement of boundaries need to be considered.
We solve the minimisation problem via bipartite matching with the following cost matrix, here illustrating the case $w=1$, where pairs at distance greater than $w$ are assigned infinite cost to prevent matching: 
\begin{align}
    \begin{split}
        C^1&(A, B) = \\
        &\begin{pmatrix}
            (A_0 - B_0)^2 & (A_0 - B_1)^2 & \infty        & \infty                & \ldots                \\
            (A_1 - B_0)^2 & (A_1 - B_1)^2 & (A_1 - B_2)^2 & \infty                & \ldots                \\
            \infty        & (A_2 - B_1)^2 & (A_2 - B_2)^2 & (A_2 - B_3)^2         & \ldots                \\
            \vdots        &               & \ddots        &                       & \vdots                \\
            \ldots        &               &               & (A_{T-1} - B_{T-2})^2 & (A_{T-1} - B_{T-1})^2 \\
        \end{pmatrix}
    \end{split}
\end{align}

For details of the bipartite matching algorithm used to solve the minimisation problem, see \cite{cormen2022introduction}. The metric is locally temporally invariant in the sense that within a time-window $w$, we have relaxed the assumption that simulations must be meaningfully ordered or aligned with respect to observations. We note that local invariance properties are not preserved under composition of permutations --- \emph{i.e.} $\pi_{w_1}(\pi_{w_2} (A,B)) \neq \pi_{w_1+w_2}(A,B)$, and that the metric $\Lc^w$ is defined according to the single locally-constrained permutation that minimises error as indicated in Eq. \ref{eq:2}.

\subsection{Bayesian Model Averaging}
Given an ensemble of $K$ plausible models $M_1,...,M_K$ predicting a quantity $y$, and training data $y_T$, Bayesian Model Averaging (BMA) provides a method of conditioning on the entire ensemble of models rather than selecting a single `best' model. 
Here, following an established BMA approach for combining weather forecast models (\cite{raftery2005using}), the predictive distribution for $y$ is given by $p(y) = \sum_{k=1}^K p(y|M_k)p(M_k|y_T)$, where $p(y|M_k)$ is the predictive distribution of an individual model $M_k$, and $p(M_k|y_T)$ is the posterior probability of $M_k$ given the training data $y_T$. 
The BMA prediction is then a weighted average of individual model predictions with weights given by the posterior probability of each model, where $\sum_{k=1}^K p(M_k|y_T)=1$.
\section[Data]{Data} \label{section:data}

Daily mean surface temperature simulations from the historical experiment of five GCMs from the latest phase of the Coupled Model Intercomparison Project (CMIP6) were used to demonstrate the method presented here: GFDL-ESM4, IPSL-CM6A-LR, MPI-ESM1-2-HR, MRI-ESM2-0 and UKESM1-0-LL.
For details of the model variants used, see Appendix \ref{appendixA}. 
These models were selected for the Inter-Sectoral Impacts Model Intercomparison Project (ISI-MIP), meeting criteria of structural independence, process representation, and historical simulation for a range of tasks. 
The subset was also found to span the range of climate sensitivity to atmospheric forcing exhibited in CMIP6 (\cite{lange2021isimip3}). 
ERA5, a high-resolution global gridded observational reanalysis, was used as a reference dataset (\cite{hersbach2020era5}). 
ERA5 hourly surface temperatures were resampled to provide daily mean temperatures.

A daily mean temperature time-series for the grid-cell containing Nairobi, Kenya was selected from each GCM and the ERA5 reference dataset.
Studies have indicated increasing heat stress risk in East African cities in recent decades (\cite{li2021spatial}); persistent CMIP model biases in simulating climate features in the region have also been noted (\cite{ongoma2018projected}), making understanding of model uncertainty in this region important.
Data were split into a training period of 1979-01-01 to 1996-12-31 and testing period of 1997-01-01 to 2014-12-31.
For each model, a simple mean-shift bias correction\footnote{Bias correction is not the focus of this work --- for a critical discussion of bias correction of systematic errors in post-processing climate model outputs for impact assessment, see \cite{ehret2012hess}.} was applied by calculating the mean error of simulated temperatures relative to ERA5 for the testing period, and subtracting this error from all model data.
For the additional experiments described in Section \ref{section:results}, the same approach was used to select GCM and ERA5 reference data for eight other cities: Paris, Chicago, Sydney, Tokyo, Kolkata, Kinshasa, Shenzhen and Santo Domingo.

A repository containing code to download GCM data, demonstrate the locally time-invariant permutation method and reproduce the results presented here is made available and can be accessed online. 

\section{Results} \label{section:results}

\begin{table}[t]
    \caption{Results from six multi-model ensemble methods for Nairobi, evaluated against ERA5 reference data. For each method, predicted number of extreme heat days \emph{n} in the train and test periods, and \emph{RMSE} for daily mean temperature predictions for these extreme heat days is shown. The locally time-invariant skill $\Lc^{15}$ for predicted temperature for extreme heat days in the test period is also shown.}
    \label{table:results}
    \centering
    \begin{tabular}{llllll}
        \toprule
        Ensemble method                & \emph{n} (train) & \emph{n} (test) & \emph{RMSE} (train) & \emph{RMSE} (test) & $\Lc^{15}$ (test) \\
        \midrule
        BMA ($\pi_{3}$)                & 721              & 967             & 1.17                & 1.31               & 0.80   \\
        BMA ($\pi_{15}$)               & 707              & 994             & 1.10                & 1.29               & 0.79   \\
        BMA ($\pi_{30}$)               & 672              & 963             & 1.08                & 1.31               & 0.83   \\
        BMA                            & 632              & 870             & 1.28                & 1.32               & 0.84   \\
        BMA (threshold)                & 726              & 973             & 1.24                & 1.31               & 0.81   \\
        MMM                            & 561              & 891             & 1.40                & 1.36               & 0.86   \\
        \midrule
        ERA5                           & 657              & 1162            &                     &                    &    \\
        \bottomrule
    \end{tabular}
\end{table}

The permutation-based method for model evaluation introduced in Section \ref{section:methodology} was demonstrated to derive multi-model ensemble predictions of daily mean temperature in nine cities.
For the training period, a permutation $\pi_w$ was applied to each simulated time-series with reference to ERA5.
Bayesian Model Averaging (BMA) was then applied to derive individual model weights and the expected value of the weighted BMA predictive distribution.
This method was tested for $w$ = 3, 15 and 31 (corresponding to matching intervals of 7, 31 and 61 days). 
These three permutation-based BMA methods are denoted BMA ($\pi_3$), BMA ($\pi_{15}$) and BMA ($\pi_{30}$) in Table \ref{table:results}.

\begin{figure}[h!]
     \centering
     \begin{subfigure}[b]{0.46\textwidth}
         \centering
         \includegraphics[width=\textwidth]{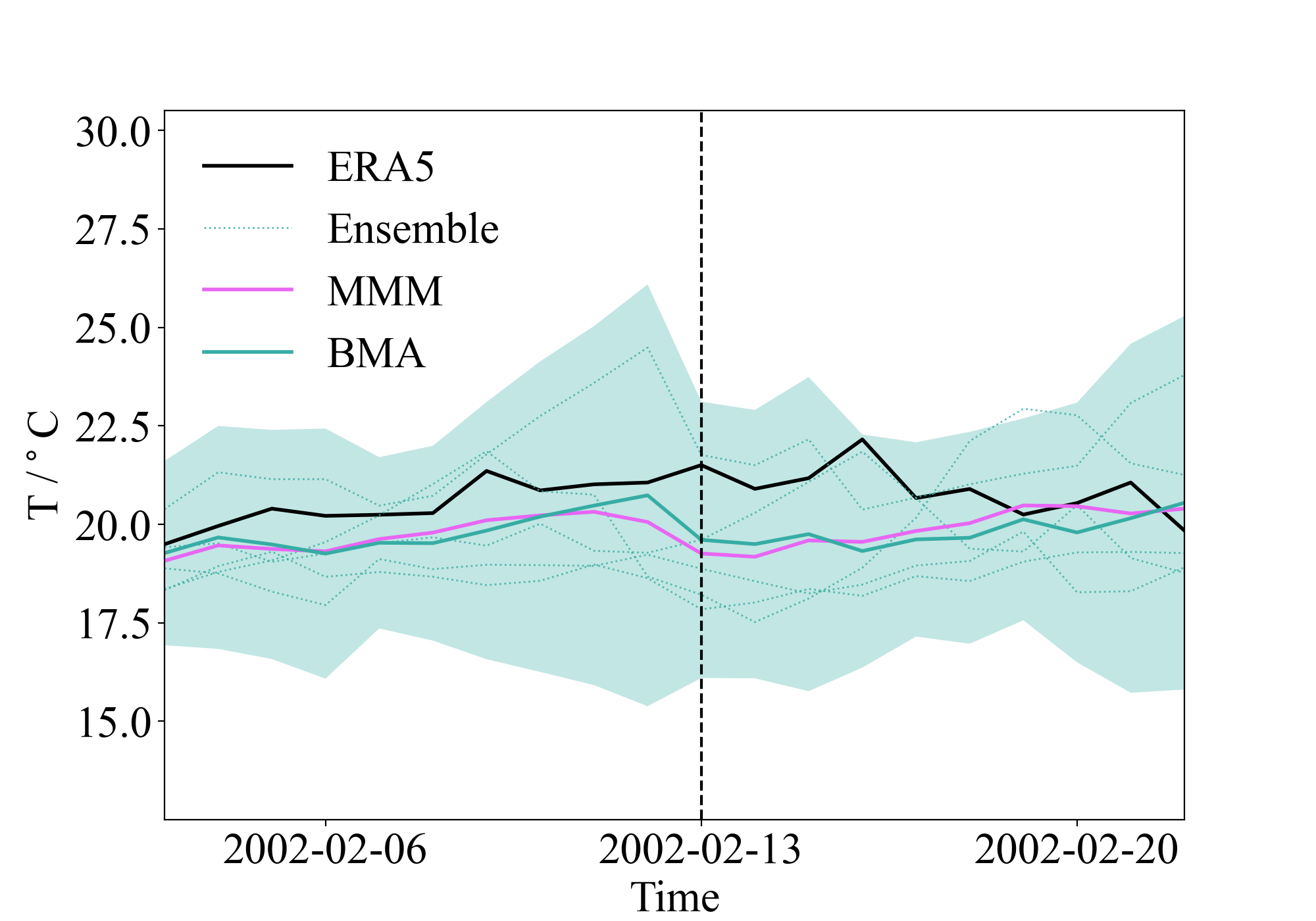}
         \caption{}
         \label{fig:series_bma}
     \end{subfigure}
     \hfill
     \begin{subfigure}[b]{0.46\textwidth}
         \centering
         \includegraphics[width=\textwidth]{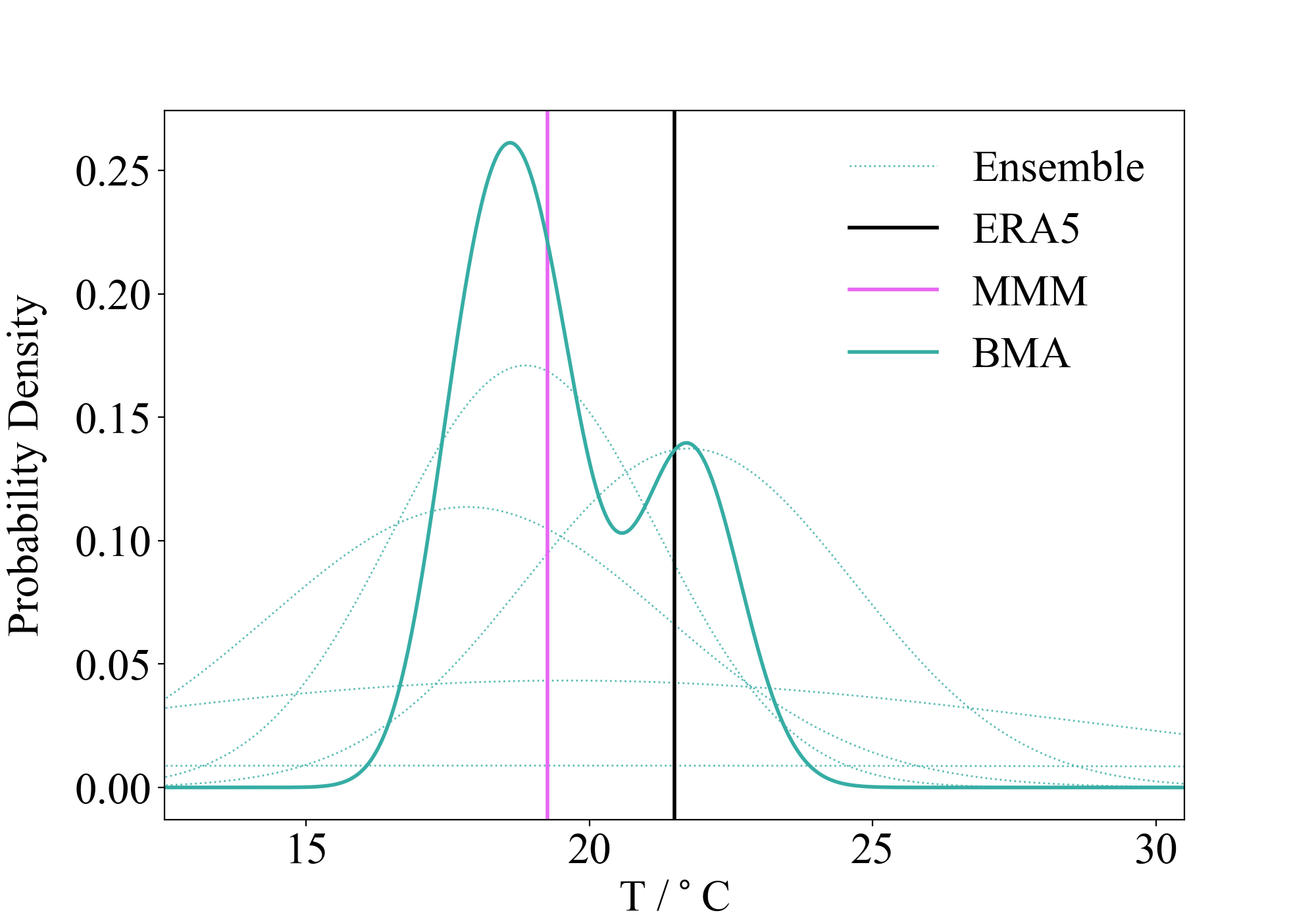}
         \caption{}
         \label{fig:dist_bma}
     \end{subfigure}\\
     \vspace{-0.5\baselineskip}
     \begin{subfigure}[b]{0.46\textwidth}
         \centering
         \includegraphics[width=\textwidth]{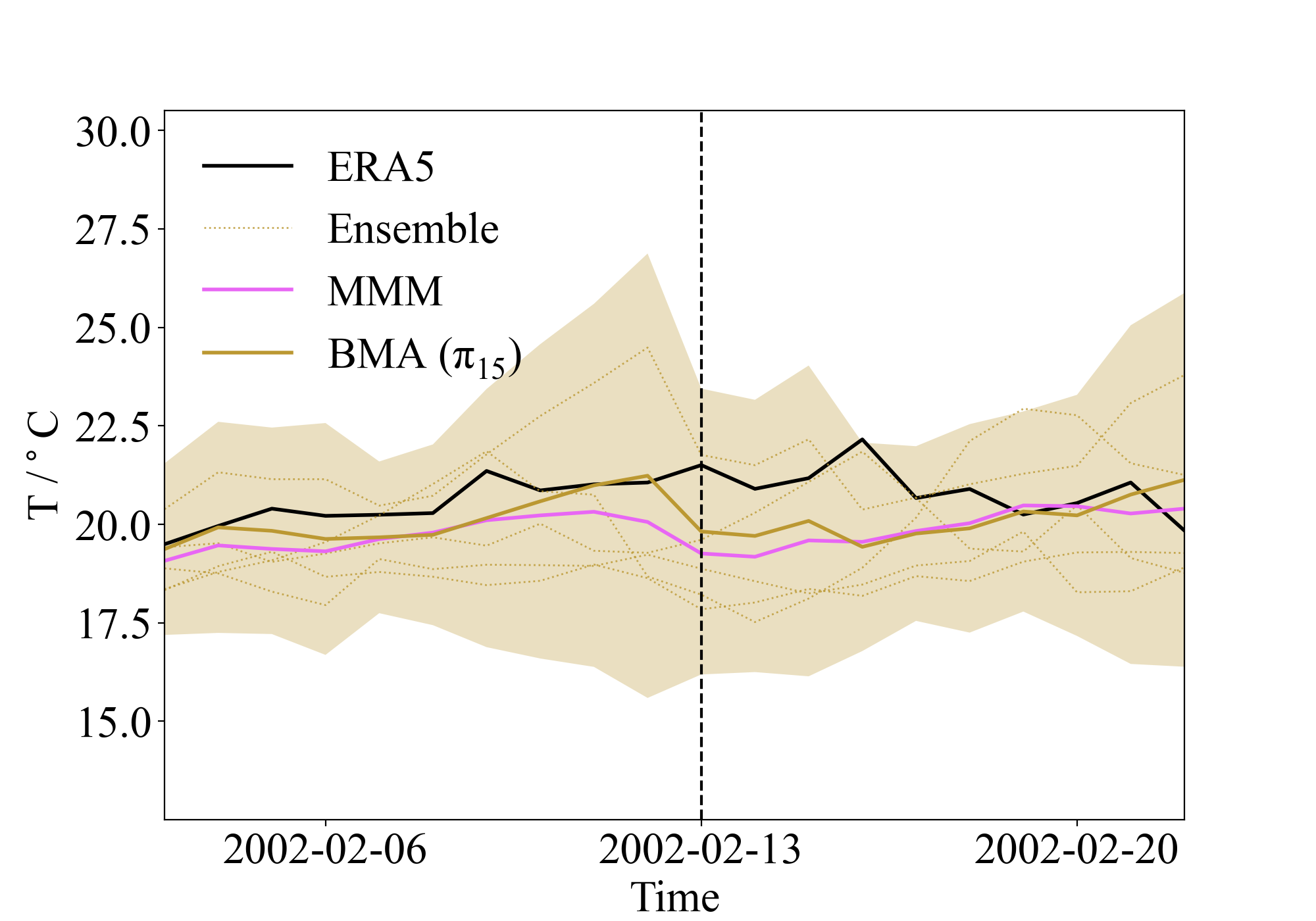}
         \caption{}
         \label{fig:series_bma15}
     \end{subfigure}
     \hfill
     \begin{subfigure}[b]{0.46\textwidth}
         \centering
         \includegraphics[width=\textwidth]{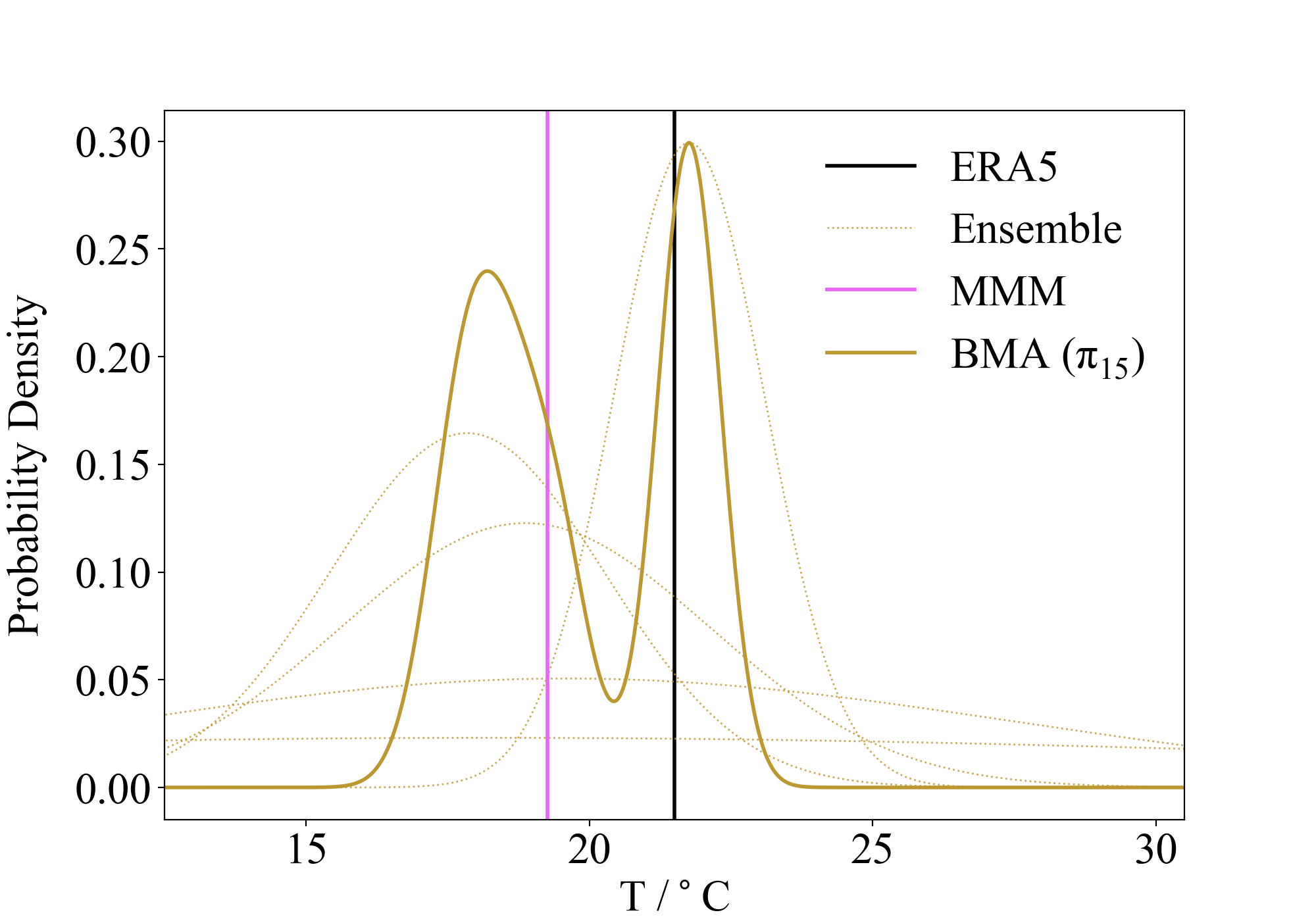}
         \caption{}
         \label{fig:dist_bma15}
     \end{subfigure}
        \caption{
        \textbf{\emph{Left:}} Sample daily mean temperature time-series from Nairobi for a period where observed daily average temperatures exceed historical 90\ts{th} quantile threshold for several consecutive days, showing individual ensemble members, ERA5 reference, MMM baseline, BMA (Figure \ref{fig:series_bma}) and BMA ($\pi_{15}$) (Figure \ref{fig:series_bma15}) predictions. Shaded region indicates $\pm$2 standard deviations from BMA predictions. Dotted vertical line indicates date of cross-section shown right. \textbf{\emph\textbf{Right:}} Cross-section of BMA (Figure \ref{fig:dist_bma}) and BMA ($\pi_{15}$) (Figure \ref{fig:dist_bma15}) predictive distributions and individual BMA-weighted ensemble members for one day. In this example, BMA ($\pi_{15}$) has assigned greater weight to a model that predicted higher temperature}
        \label{fig:BMA_demo}
\end{figure}

Three baseline methods were also implemented for comparison: a simple Multi-Model Mean (MMM) approach, standard BMA without permutation, and a modified BMA approach (denoted BMA (threshold)) where only simulation of observed extreme heat days was considered when calculating the model weights. 
The results from each of these six methods for the city of Nairobi are shown in Table~\ref{table:results}. 
The results in Table~\ref{table:results} show the valuation according to the predicted number of extreme heat days and RMSE in predicted mean temperature for these days. 
Additionally, the locally time invariant skill metric $\Lc^{15}$ for the extreme heat days in the test period is shown.

\begin{figure}[t]
     \centering
     \hspace{-0.01\textwidth}
     \begin{subfigure}[b]{0.04\textwidth}
         \includegraphics[width=\textwidth, angle=90]{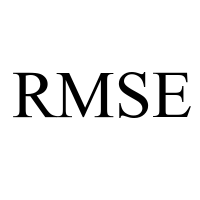}
         \vspace{5\textwidth}
     \end{subfigure}
     \hspace{-0.01\textwidth}
     \begin{subfigure}[b]{0.4\textwidth}
         \centering
         \includegraphics[width=\textwidth]{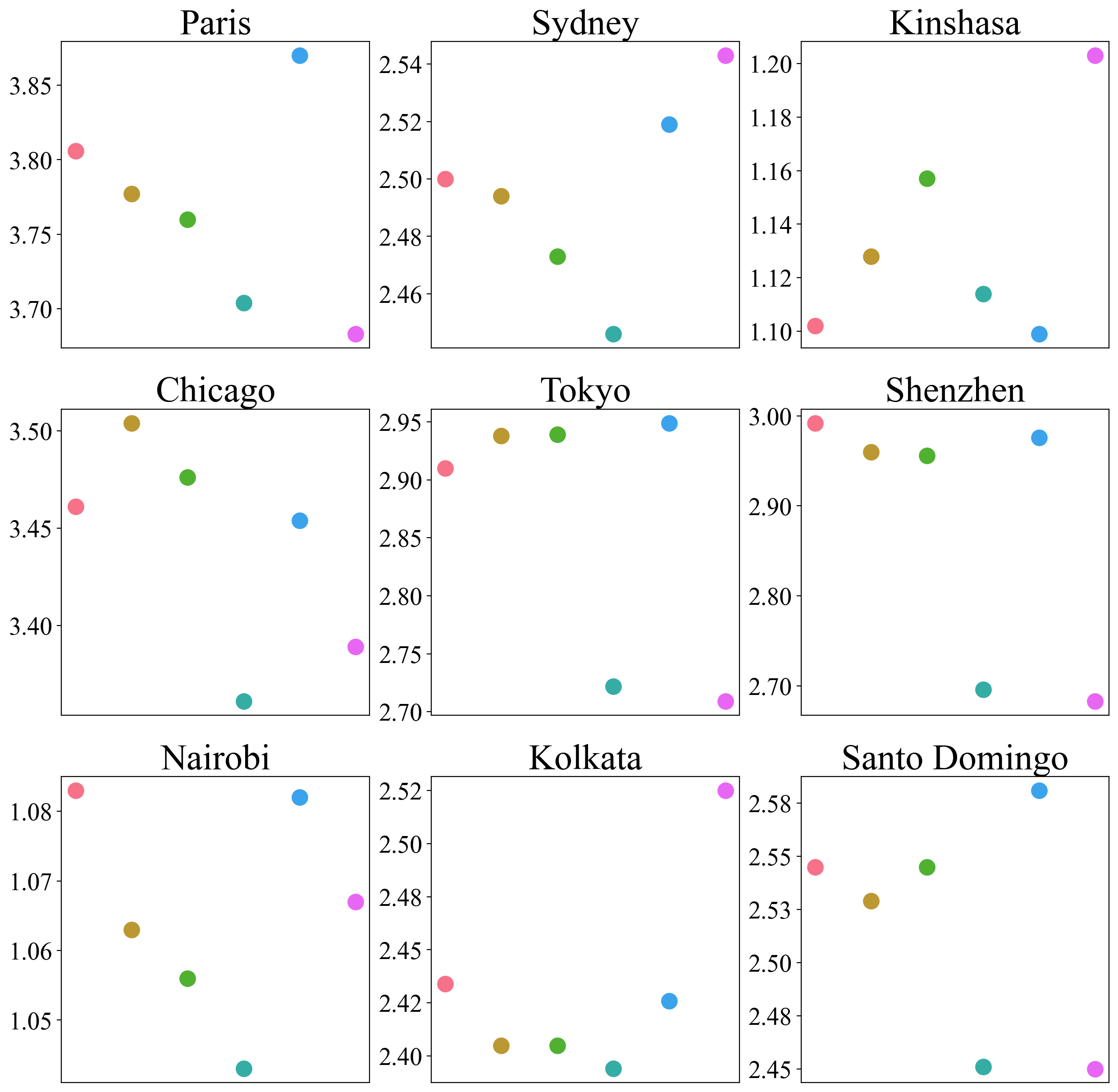}
         \caption{RMSE in daily mean temperature for all days}
         \label{fig:all_cities_rmse}
     \end{subfigure}
     \hfill
     \begin{subfigure}[b]{0.4\textwidth}
         \centering
         \includegraphics[width=\textwidth]{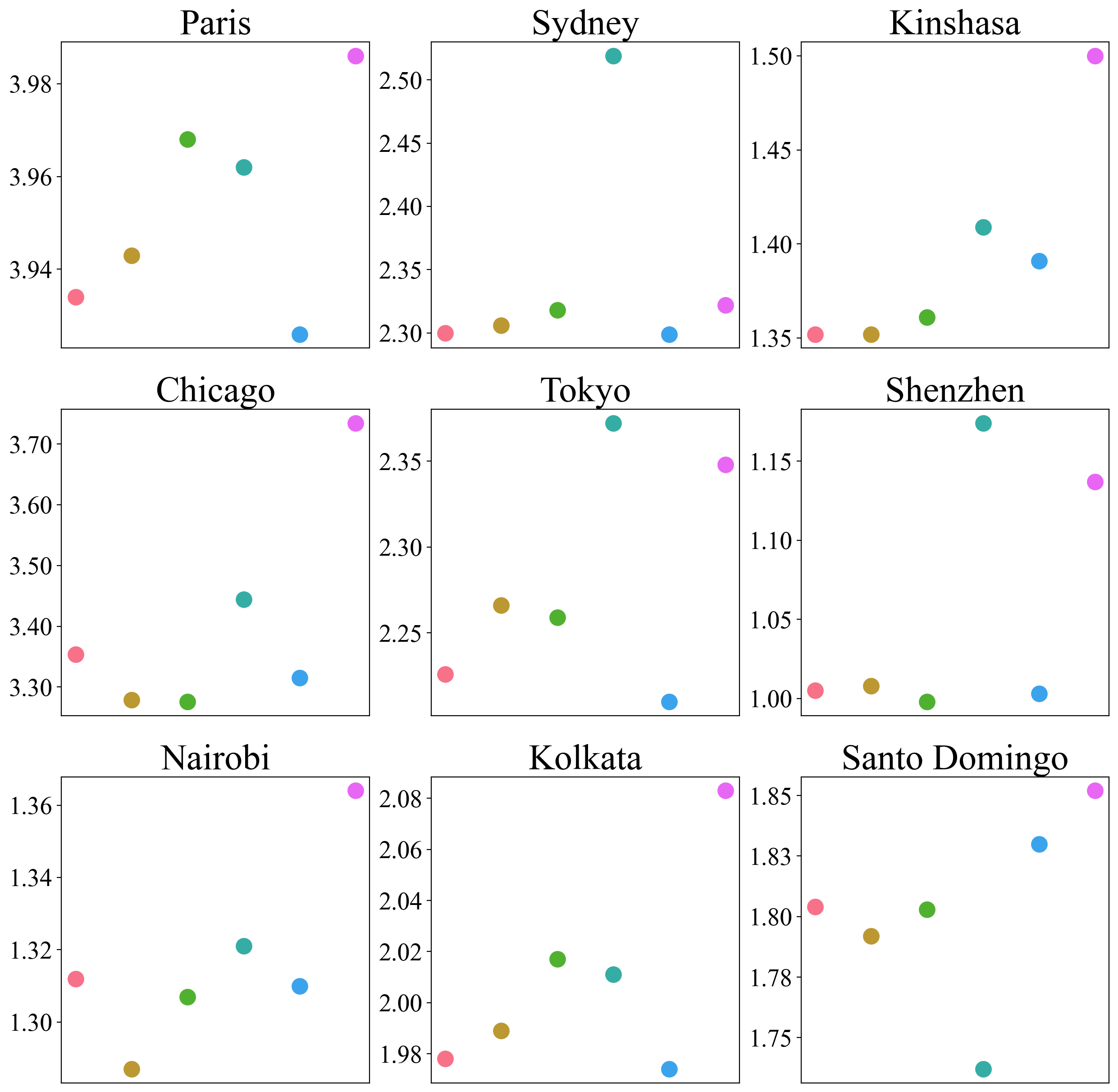}
         \caption{RMSE in daily mean temperature for extreme heat days}
         \label{fig:all_cities_rmse_q90}
     \end{subfigure}
     \begin{subfigure}[b]{0.13\textwidth}
         \includegraphics[width=\textwidth]{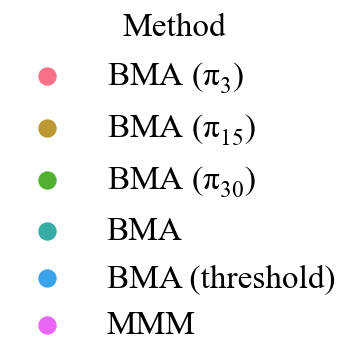}
         \vspace{1.3\textwidth}
     \end{subfigure}
     \caption{Evaluation of six multi-model ensemble methods for experiments across nine cities, showing \ref{fig:all_cities_rmse}: RMSE for predicting daily average temperature for all days; and \ref{fig:all_cities_rmse_q90}: RMSE for predicting daily average temperature for extreme heat days.}
 \label{fig:rmse}
\end{figure}

Figures \ref{fig:series_bma} and \ref{fig:series_bma15} show a short sample time-series of the reference and ensemble simulation data from the test period, showing the predictions given by MMM, standard BMA (\ref{fig:series_bma}) and BMA ($\pi_{15}$) (\ref{fig:series_bma15}), with a $\pm$ 2 standard deviation region shaded for the BMA methods. 
Figures \ref{fig:dist_bma} and \ref{fig:dist_bma15} show a cross-section for a single day from this time-series indicating the BMA and BMA ($\pi_{15}$) predictive distributions as a combination of the weighted ensemble members. These experiments were repeated for eight other cities.
The results from each of the six methods is compared for each city in Figure \ref{fig:rmse}.

\begin{figure}[t]
     \centering
     \hspace{-0.01\textwidth}
     \begin{subfigure}[b]{0.04\textwidth}
         \includegraphics[width=\textwidth, angle=90]{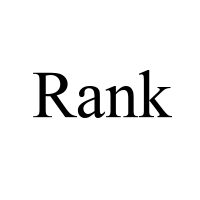}
         \vspace{0.5\textwidth}
     \end{subfigure}
     \hspace{-0.01\textwidth}
     \begin{subfigure}[]{0.31\textwidth}
         \centering
         \includegraphics[width=\textwidth]{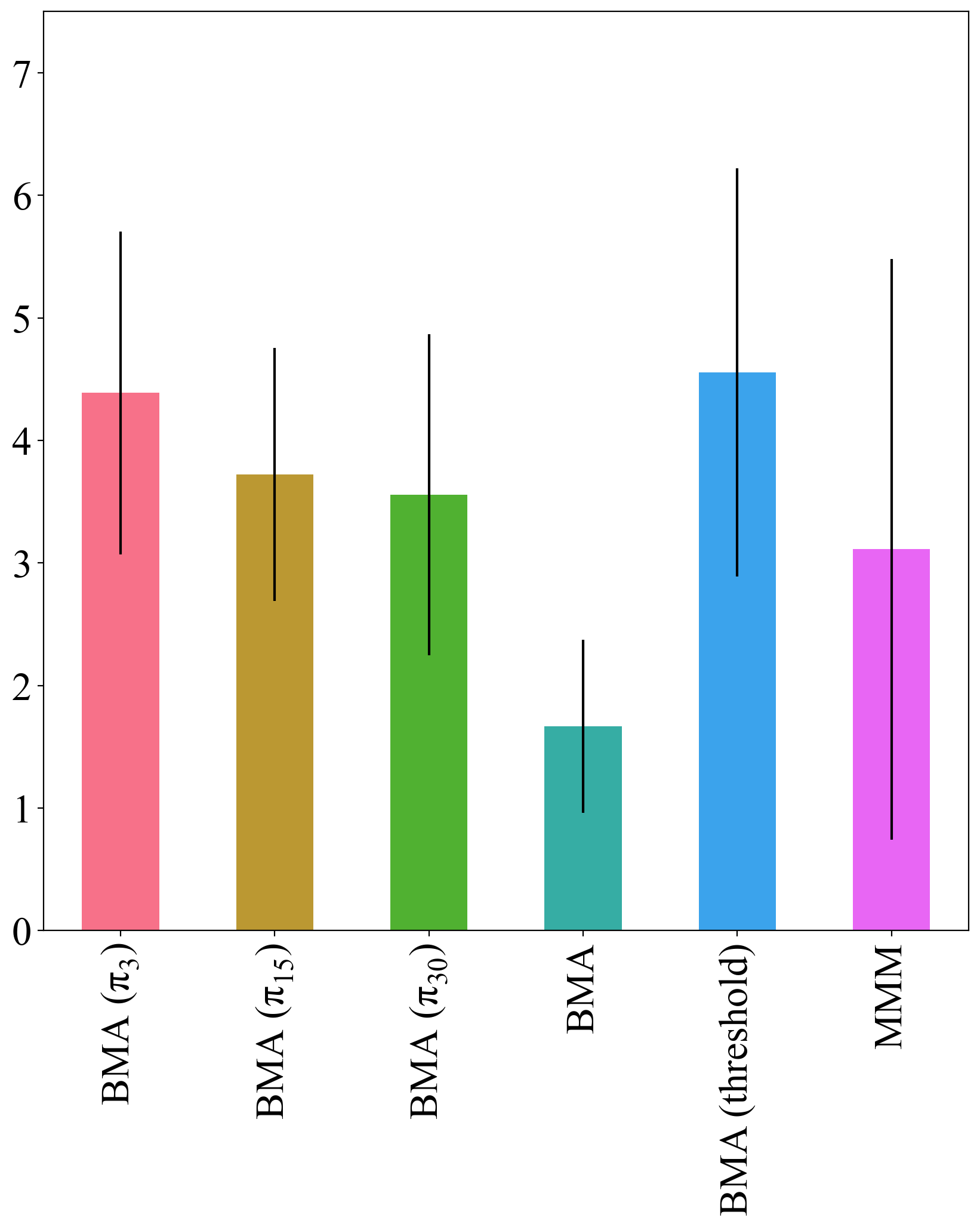}
         \caption{RMSE in daily average temperature for all days}
         \label{fig:ranks_all}
     \end{subfigure}
     \begin{subfigure}[]{0.31\textwidth}
         \centering
         \includegraphics[width=\textwidth]{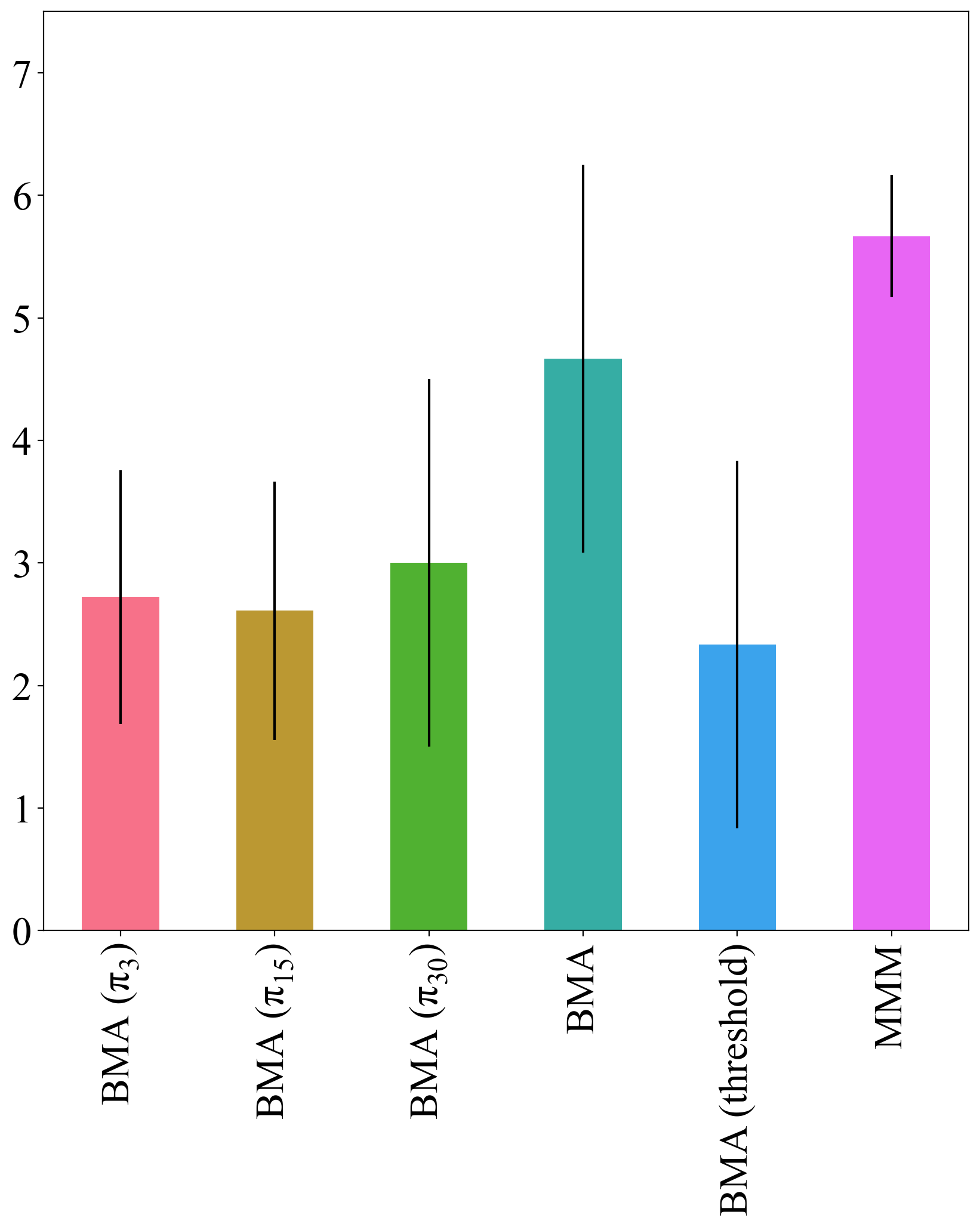}
         \caption{RMSE in daily average temperature for extreme heat days}
         \label{fig:ranks_q90}
     \end{subfigure}
     \begin{subfigure}[]{0.31\textwidth}
         \centering
         \includegraphics[width=\textwidth]{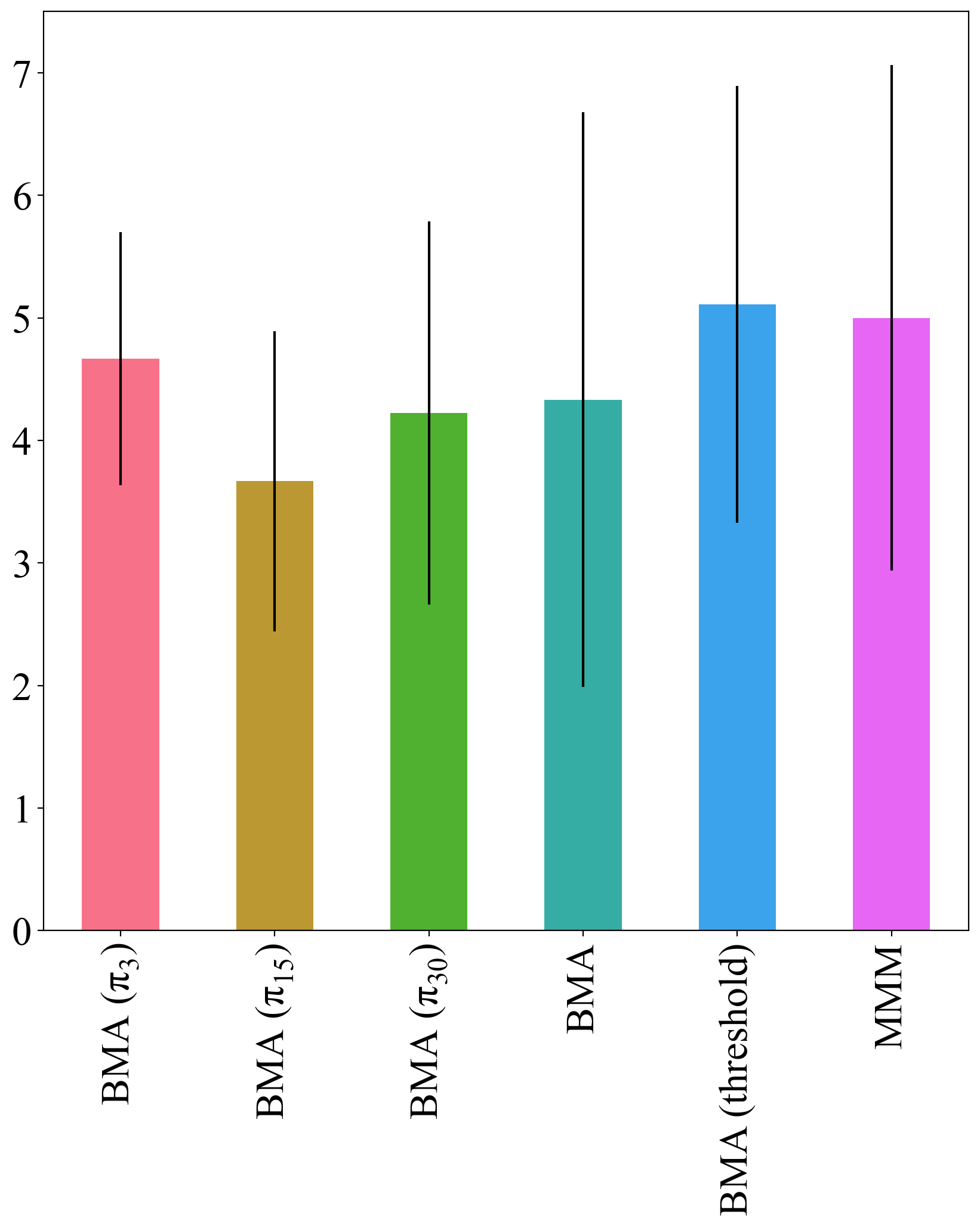}
         \caption{Error in number of extreme heat days}
         \label{fig:ranks_n}
     \end{subfigure}
     \caption{Summary of rankings of six multi-model ensemble methods for experiments across nine cities, ranked by \ref{fig:ranks_all}: RMSE in predicting daily average temperature for all days; \ref{fig:ranks_q90}: RMSE in predicting daily average temperature for extreme heat days; and \ref{fig:ranks_n}: absolute error for predicting number of extreme heat days. In each case the best-performing method is rank $1$}
     \label{fig:ranks}
\end{figure}

\begin{figure}[t]
    \hspace{0.14\textwidth}
    \begin{subfigure}[b]{0.04\textwidth}
        \includegraphics[width=\textwidth, angle=90]{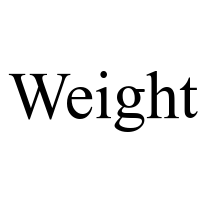}
        \vspace{7\textwidth}
    \end{subfigure}
    \hspace{-0.01\textwidth}
    \includegraphics[width=0.76\textwidth]{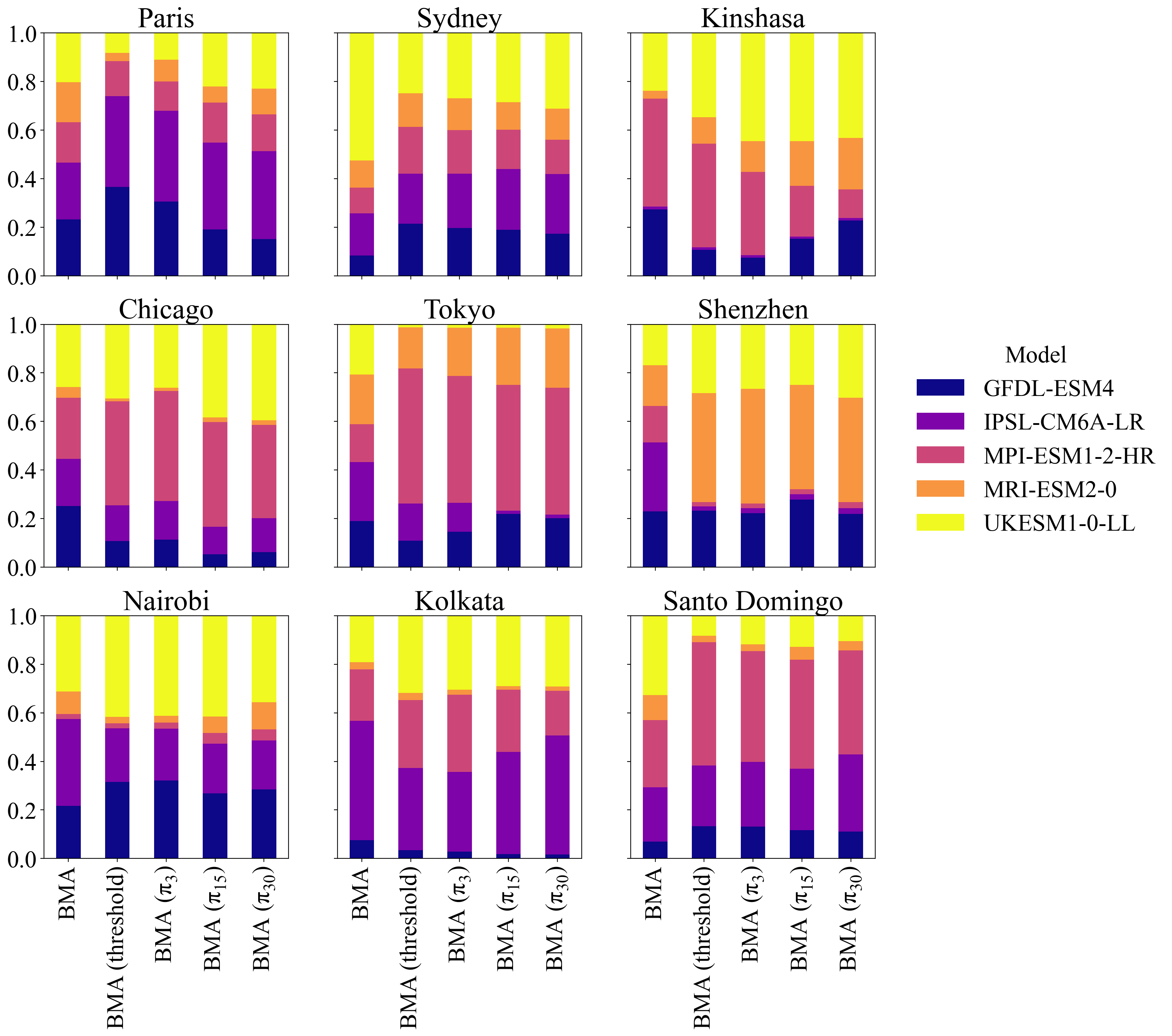}
    \caption{Climate model weights calculated from five BMA methods for experiments from nine cities}
    \label{fig:weights}
\end{figure}

\begin{figure}[t]
     \hspace{0.12\textwidth}
     \begin{subfigure}[b]{0.04\textwidth}
         \includegraphics[width=\textwidth, angle=90]{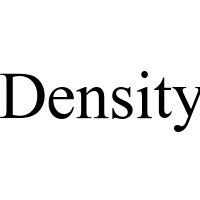}
         \vspace{5.5\textwidth}
     \end{subfigure}
     \hspace{-0.03\textwidth}
     \begin{subfigure}[b]{0.63\textwidth}
        \centering
        \begin{subfigure}{0.3\textwidth}
            \centering
            \includegraphics[width=\linewidth]{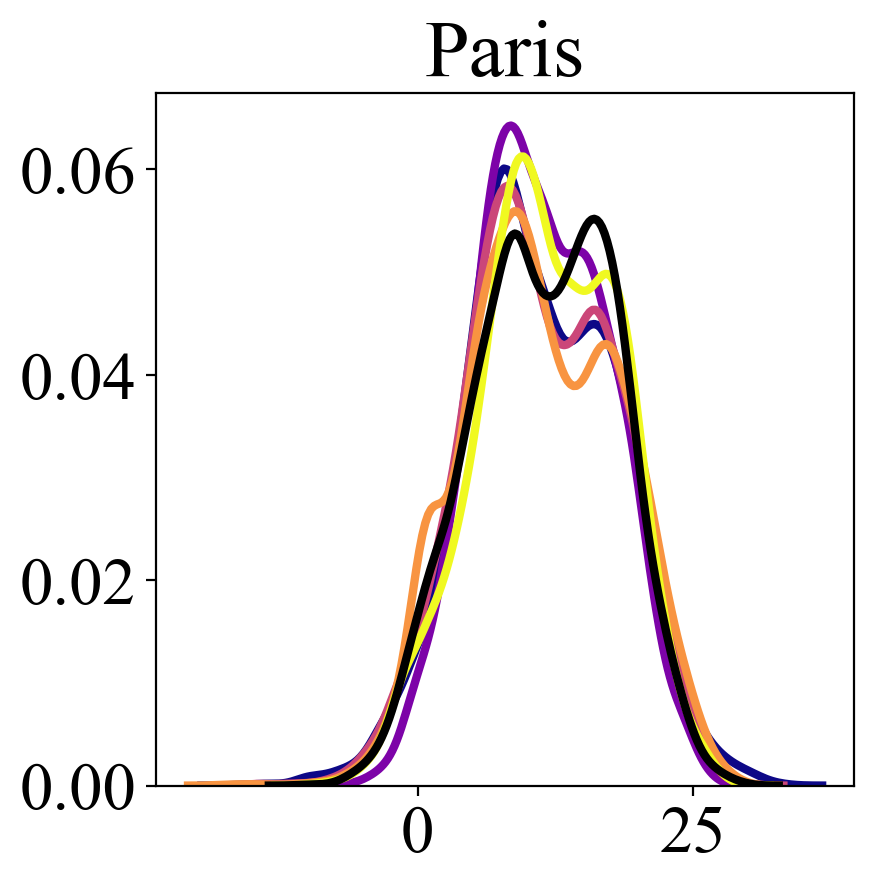}
        \end{subfigure}
        \begin{subfigure}{.3\textwidth}
            \centering
            \includegraphics[width=\linewidth]{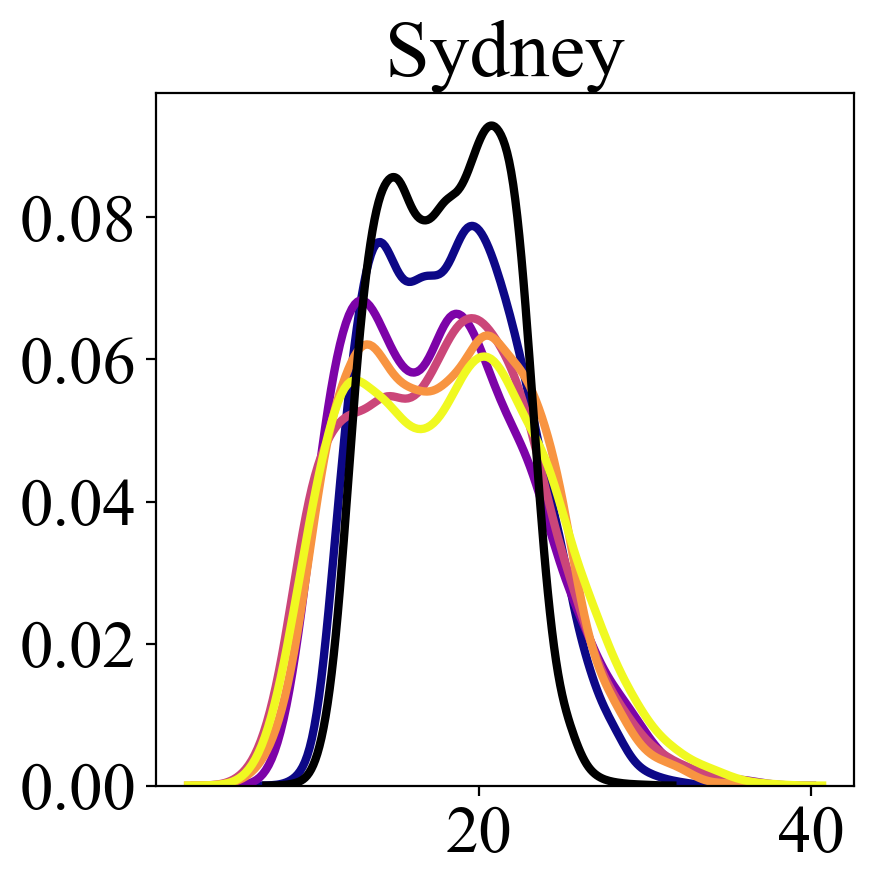}
        \end{subfigure}
        \begin{subfigure}{.3\textwidth}
            \centering
            \includegraphics[width=\linewidth]{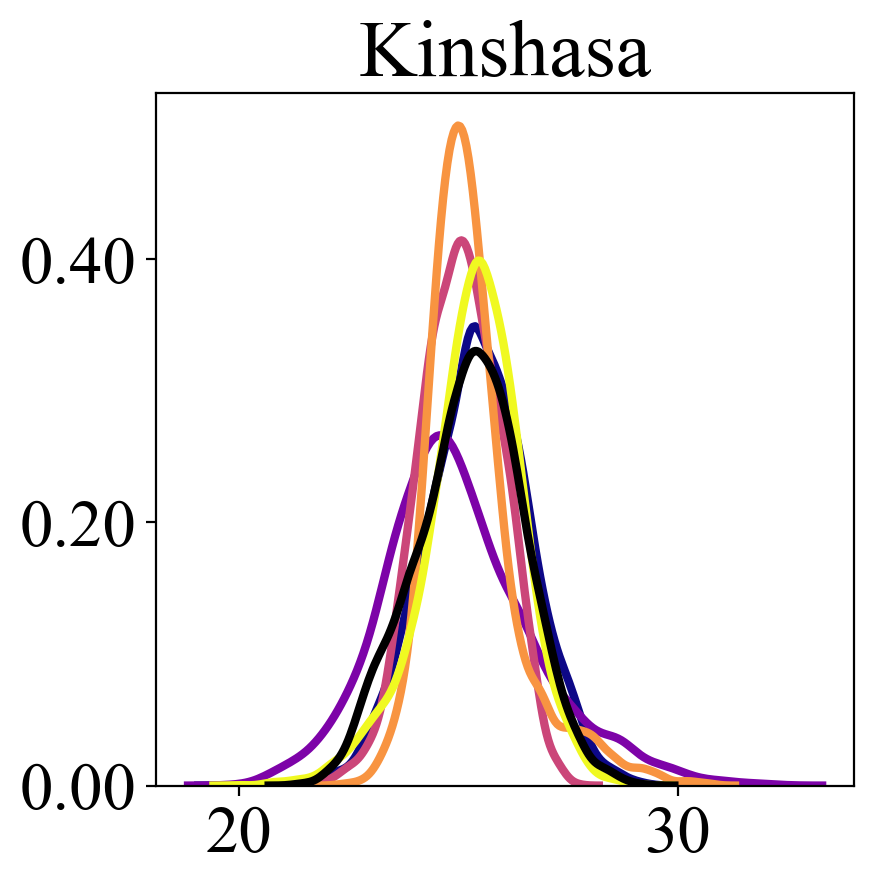}
        \end{subfigure}\\
        \vspace{-1.07\baselineskip}
        \begin{subfigure}{0.3\textwidth}
            \centering
            \includegraphics[width=\linewidth]{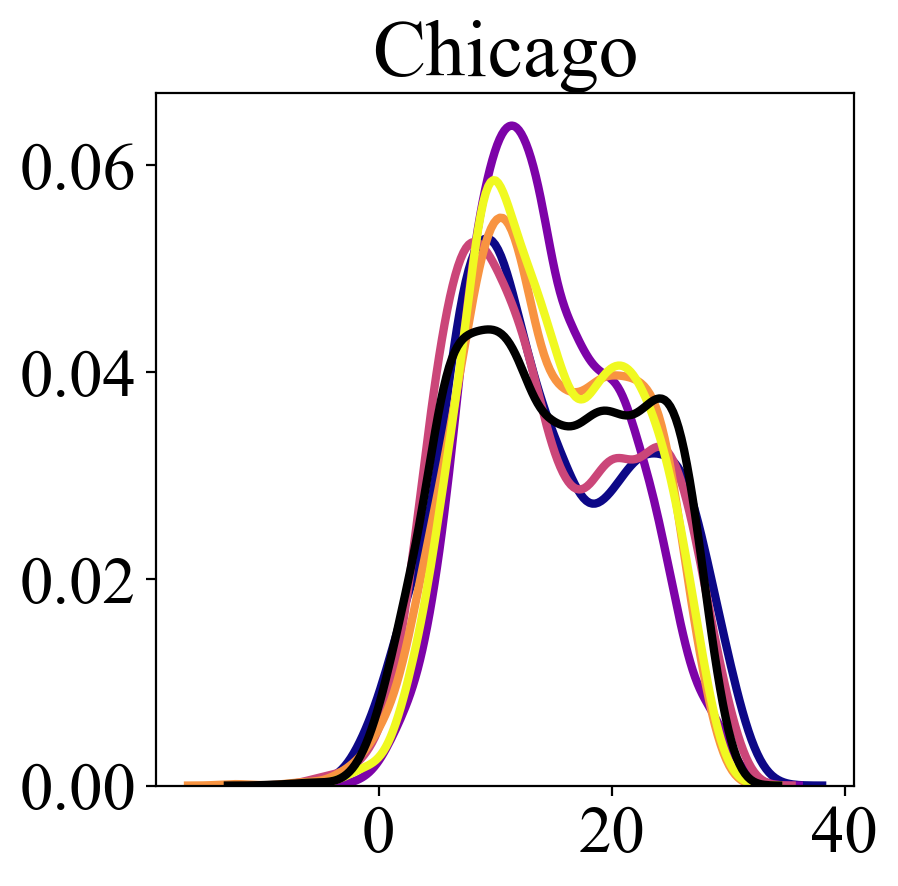}
        \end{subfigure}
        \begin{subfigure}{.3\textwidth}
            \centering
            \includegraphics[width=\linewidth]{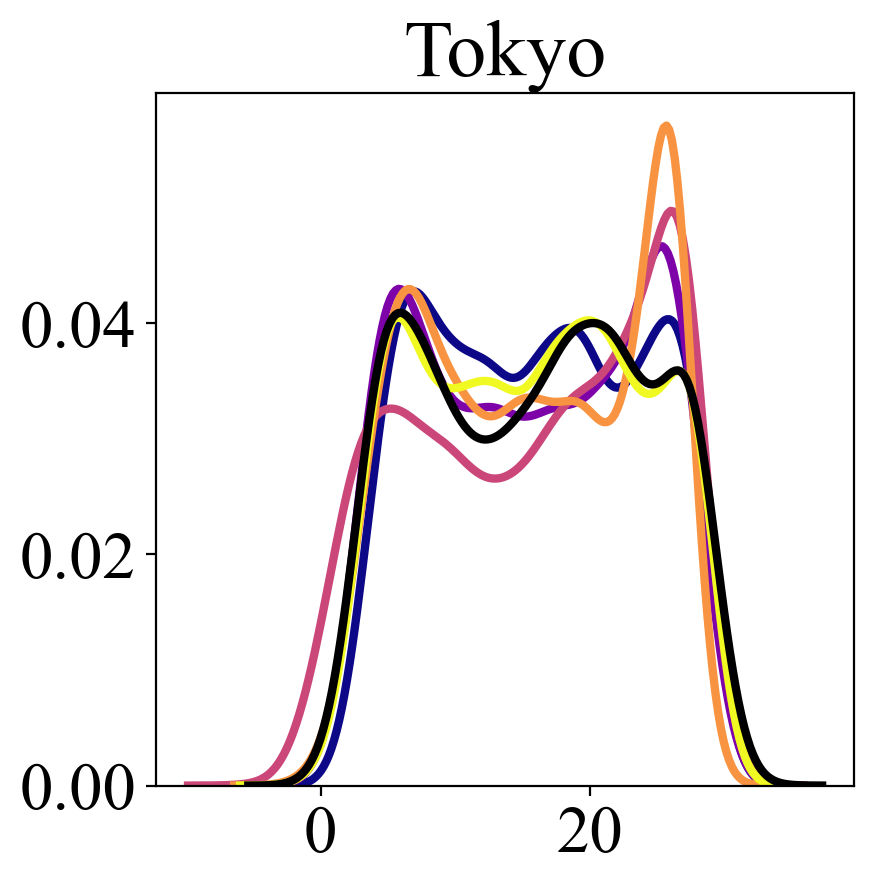}
        \end{subfigure}
        \begin{subfigure}{.3\textwidth}
            \centering
            \includegraphics[width=\linewidth]{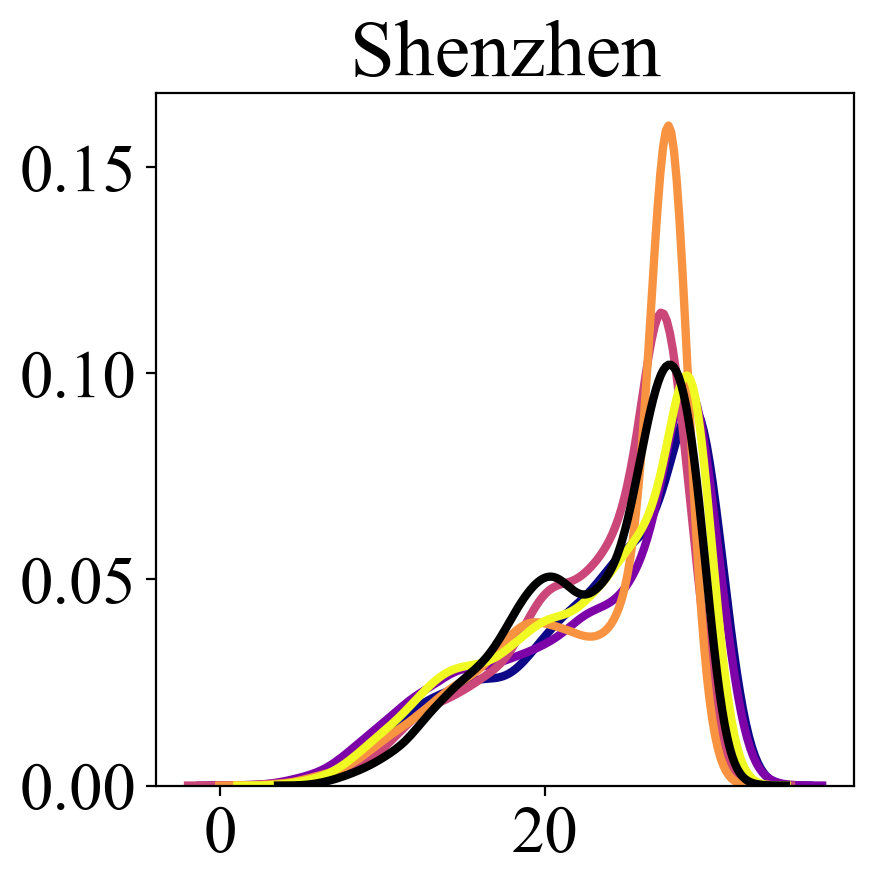}
        \end{subfigure}\\
        \vspace{-1.07\baselineskip}
        \begin{subfigure}{0.3\textwidth}
            \centering
            \includegraphics[width=\linewidth]{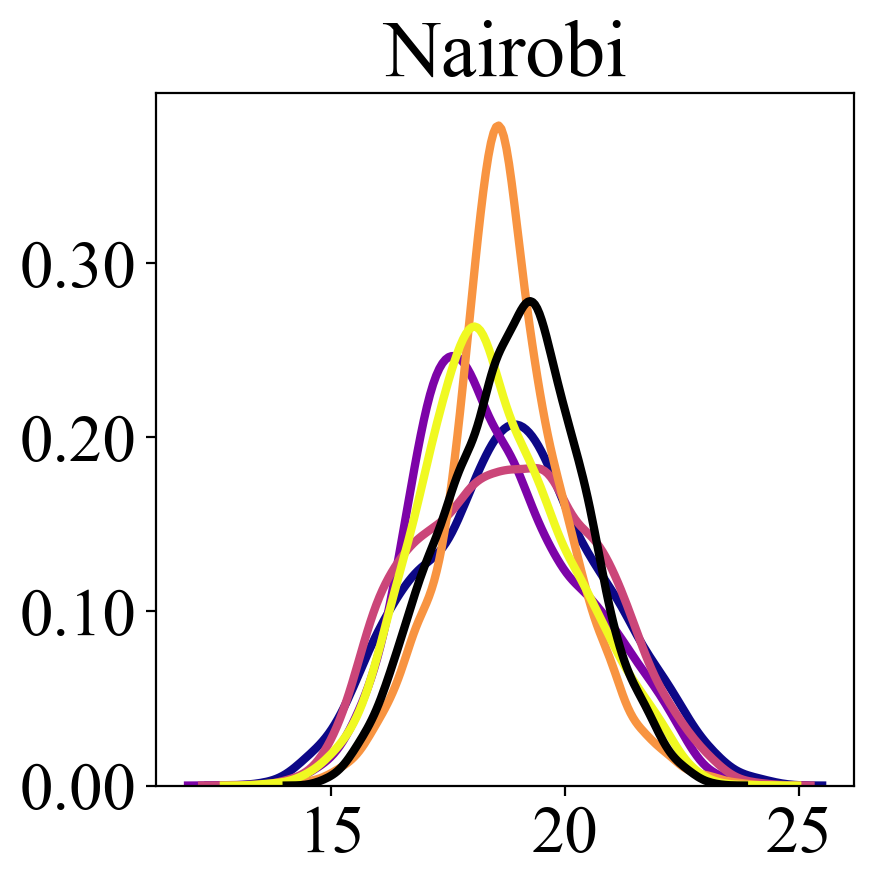}
        \end{subfigure}
        \begin{subfigure}{.3\textwidth}
            \centering
            \includegraphics[width=\linewidth]{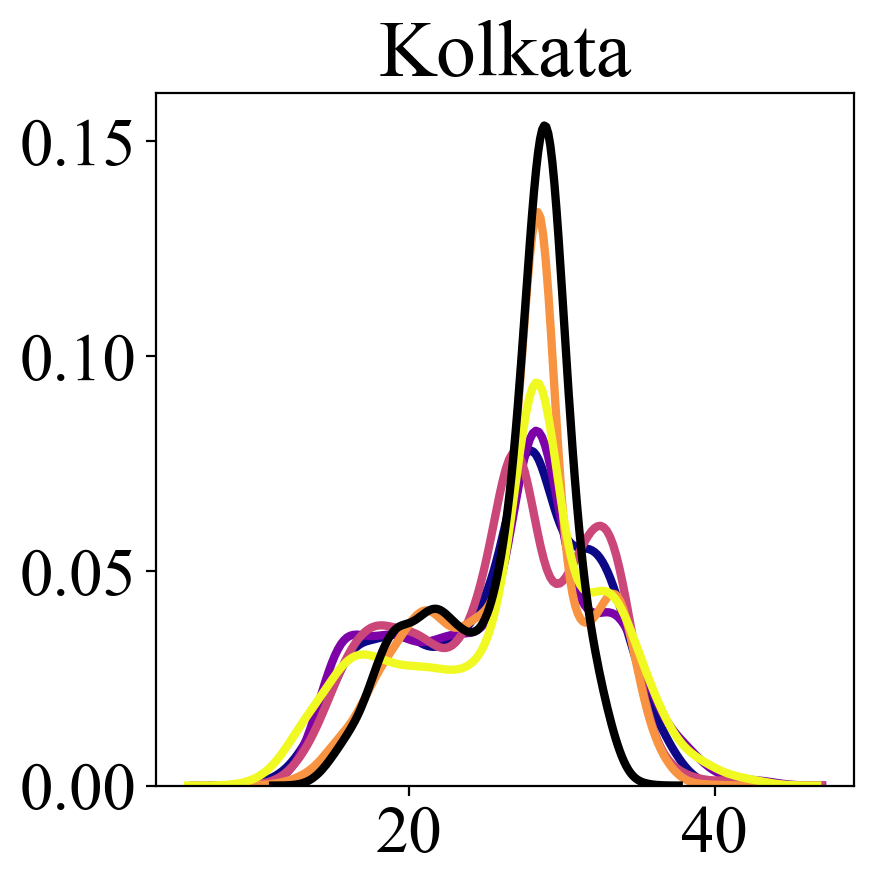}
        \end{subfigure}
        \begin{subfigure}{.3\textwidth}
            \centering
            \includegraphics[width=\linewidth]{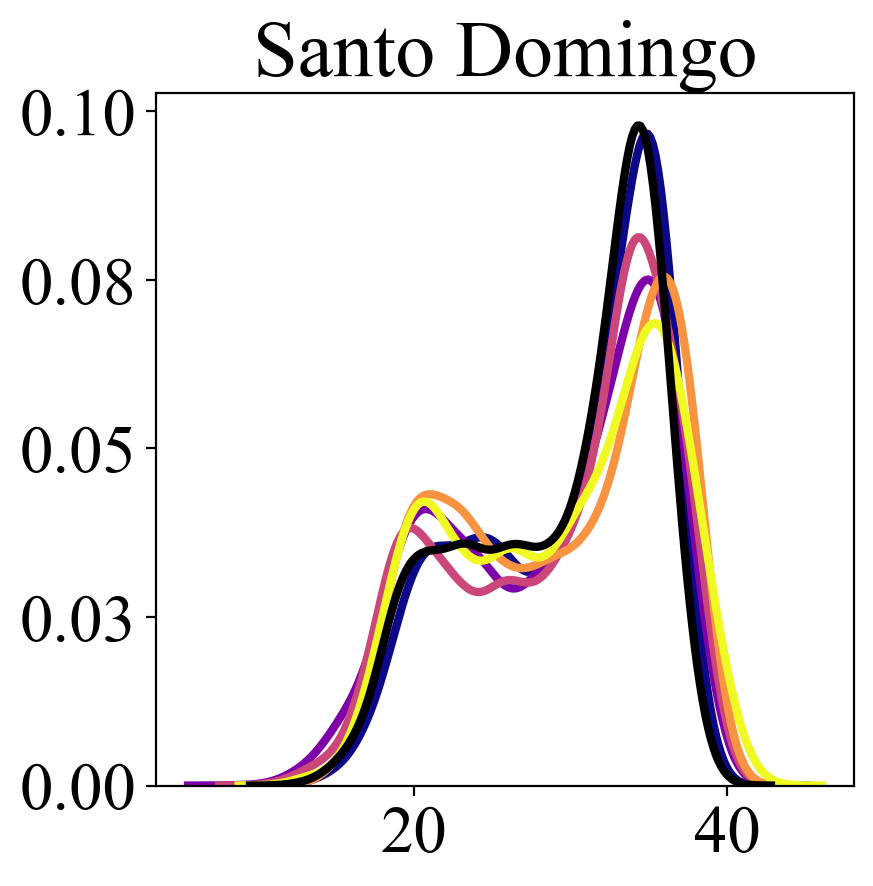}
        \end{subfigure}
    \end{subfigure}
    \hspace{-0.03\textwidth}
    \begin{subfigure}[b]{0.17\textwidth}
        \includegraphics[width=\textwidth]{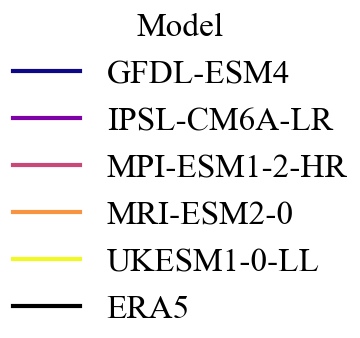}
        \vspace{\textwidth}
    \end{subfigure}\\         
    \centering
    \begin{subfigure}[t]{0.04\textwidth}
        \vspace{-1.5\textwidth}
        \hspace{-0.7\textwidth}
        \includegraphics[width=\textwidth]{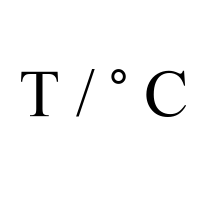}
    \end{subfigure}
    \caption{Distributions of daily average temperature for test period from each GCM simulation and ERA5 reference for nine cities}
    \label{fig:distributions}
\end{figure}

A summary plot ranking the best-performing method across these experiments is shown in Figure \ref{fig:ranks}. The individual model weights calculated by each of the six methods for each city are shown in Figure \ref{fig:weights}. To aid interpretation of these model weights, the distribution of the data from each model alongside ERA5 reference for the training period is shown in Figure \ref{fig:distributions}. (Note that a mean-shift bias correction has been applied to the distributions as described in Section \ref{section:data}).

\section{Discussion} 

The results from the six methods applied to derive multi-model ensemble predictions of daily average temperature for Nairobi (Table \ref{table:results}) indicate that all BMA approaches outperformed the MMM baseline both in predicting the number of extreme heat days $n$ in the test period, and RMSE for these days.
BMA after applying a $\pi_{15}$ permutation was the best performing method for predicting both $n$ and RMSE of extreme heat days by a small margin. 
Experiments applying permutations across a range of window sizes $\pi_w$ indicated that the window size $w$=15 (corresponding to an allowed matching interval of 31 days) tended to perform well consistently.
Consequently, an additional evaluation of each ensemble method in terms of $\Lc^{15}$ is also shown in Table \ref{table:results}, indicating that BMA ($\pi_{15}$) performs best according to this metric.

Further experiments across eight other cities (Figure \ref{fig:rmse}) indicate that whilst the MMM and standard BMA approaches performed well in predicting RMSE for all days in the test period, the ensemble methods more tailored towards predicting extreme heat days --- BMA (threshold), BMA ($\pi_3$), BMA ($\pi_{15}$), and BMA ($\pi_{30}$) --- outperformed these baselines across all locations for predicting RMSE of extreme heat days. 
In general we note that RMSE for predicting extreme heat days decreases with $w$ up to a point, and then begins to increase for larger values (see BMA ($\pi_{30}$) in Figure \ref{fig:all_cities_rmse_q90}). 
Comprehensive experiments into the effect of window size would be required to draw stronger conclusions regarding the optimal value for a given geographical location.

The rankings of each ensemble method (Figure \ref{fig:ranks}) similarly indicate that whilst standard BMA and MMM approaches consistently performed well for predicting RMSE for all days, the permutation-based approaches and BMA (threshold), which considered only extreme heat days when assigning model weights, performed better for predicting RMSE for extreme heat days.
The permutation-based methods outranked other methods for predicting number of extreme heat days, including the BMA (threshold) approach, suggesting that the introduction of local temporal invariance before model evaluation has led to a better-informed model weighting.
These results indicate that there is a need to customise multi-model ensemble schemes for the prediction of extremes. We note that the effect sizes in the results presented here are small, and analysis of their consistency across other locations and test periods is an area for future study.

The weights assigned to individual models by the five BMA methods are shown in Figure \ref{fig:weights}.
To aid the interpretation of these weights, the distribution of each simulation and the ERA5 reference for the test period is shown in Figure \ref{fig:distributions}. 
For some cities, it is apparent that low model weights have been assigned where the simulated distribution differs significantly from the reference distribution --- see for example MRI-EM2-0 and MPI-ESM1-2-HR for Nairobi, and IPSL-CM6A-LR and MRI-ESM2-0 for Kinshasa. 

In several cities, model weights vary substantially between standard BMA and the other approaches (see Shenzhen and Tokyo), again highlighting the need to modify model weighting schemes for optimal prediction of extremes.
In general, it can be noted that whilst standard BMA assigns relatively even weightings to each model in the ensemble, the permutation-based approaches impose greater sparsity on the ensemble.
Relaxing the assumption of temporal alignment during model evaluation, therefore, allows a stronger distinction to be made regarding which models should be considered skillful for a particular location.

Repetition of these experiments using alternative realisations of each model (\emph{i.e.} a different `run' of the same climate model using the same parameters and initial conditions, simulating an alternative pathway given the inherent randomness of the climate system) yielded some variance in the assignation of model weights but broad consistency in the ranking of ensemble methods --- results for ensemble methods applied to an alternative set of model realisations for Nairobi is provided in Appendix \ref{appendixB} (see Figure \ref{fig:alt_realisations}). 
The method has been demonstrated here for daily average temperature prediction --- however, the same reasoning could also be extended to other simulated climate variables in future work.

\section{Conclusion}
We present a novel permutation-based method for the evaluation of climate model simulations that introduces local temporal invariance. This enables us to relax the assumption that simulated extremes should be temporally aligned or ordered without reducing the temporal precision of models. This evaluation method is tested within a Bayesian Model Averaging multi-model ensemble weighting scheme to derive probabilistic predictions of extreme heat days for nine cities. Our results highlight the need for model evaluation methods tailored for assessing the simulation of extremes when producing multi-model ensemble projections for impact assessment and adaptation planning. We find that incorporation of local temporal invariance during model evaluation enables a more skilful model weighting to be derived, yielding improved prediction of the number of extreme heat days and RMSE for these days compared to standard BMA. We highlight directions for future work, including advancement of the methodology presented here and approaches to tailor ensemble methods for predictions of extreme events.

\paragraph{Impact Statement} Adaptation to climate change requires predictions of how the frequency and severity of extreme events will change in the future. 
Here, we consider the occurrence of extreme heat days in cities, which pose serious societal risks including exceedance of human heat stress thresholds. 
We propose a method for combining multiple climate model simulations that optimises predictions of such extreme events, and demonstrate the advantages of this method for nine cities.

\printbibliography

\clearpage
\appendix

\section{Appendix}
\subsection{Data}\label{appendixA}

\begin{table}[h]
    \caption{CMIP6 models and realisations used to produce the results presented in Section \ref{section:results}}
    \label{table:GCM_data}
    \centering
    \begin{tabular}{lll}
        \toprule
        Model name    & Realisations      & Experiment \\
        \midrule
        GFDL-ESM4     & \texttt{r1i1p1f1} & Historical \\
        IPSL-CM6A-LR  & \texttt{r1i1p1f1} & Historical \\
        MPI-ESM1-2-HR & \texttt{r1i1p1f1} & Historical \\
        MRI-ESM2-0    & \texttt{r1i1p1f1} & Historical \\
        UKESM1-0-LL   & \texttt{r1i1p1f2} & Historical \\
        \bottomrule
    \end{tabular}
\end{table}

\subsection{Additional experiments}\label{appendixB}

\begin{figure}[H]
     \centering
     \hspace{-0.01\textwidth}
     \begin{subfigure}[b]{0.04\textwidth}
         \includegraphics[width=\textwidth, angle=90]{figures/RMSE_label.png}
         \vspace{0.8\textwidth}
     \end{subfigure}
     \hspace{-0.01\textwidth}
     \begin{subfigure}[t]{0.21\textwidth}
         \centering
         \includegraphics[width=\textwidth]{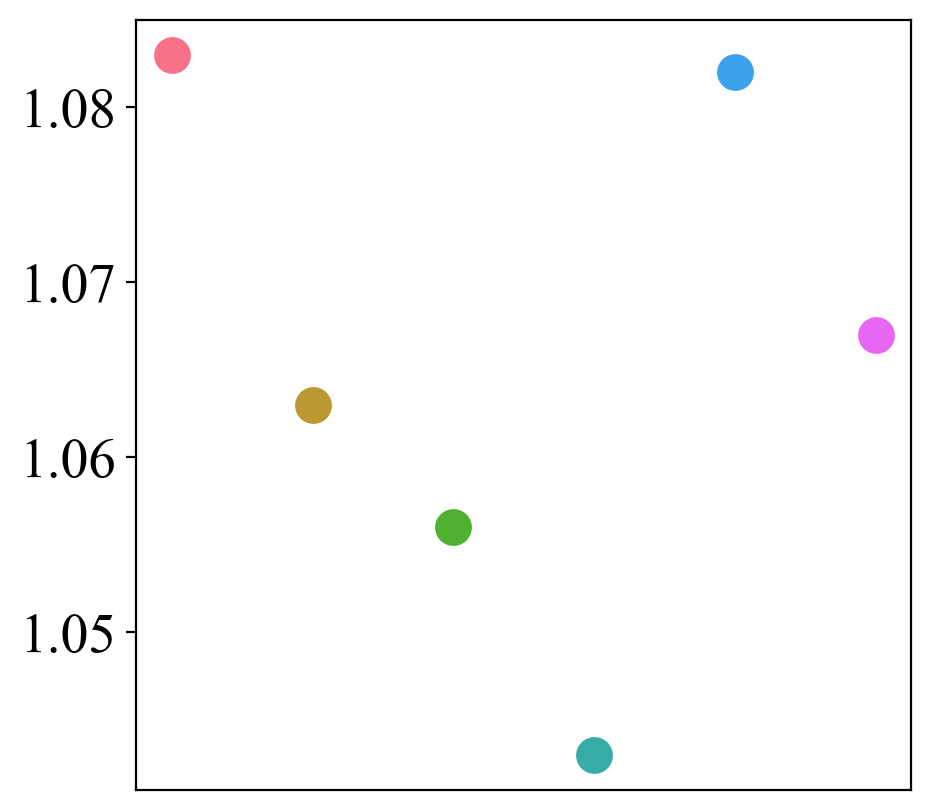}
         \caption{RMSE in daily mean temperature for all days}
         \label{fig:alt_rmse_all}
     \end{subfigure}
     \hspace{0.02\textwidth}
     \begin{subfigure}[t]{0.21\textwidth}
         \centering
         \includegraphics[width=\textwidth]{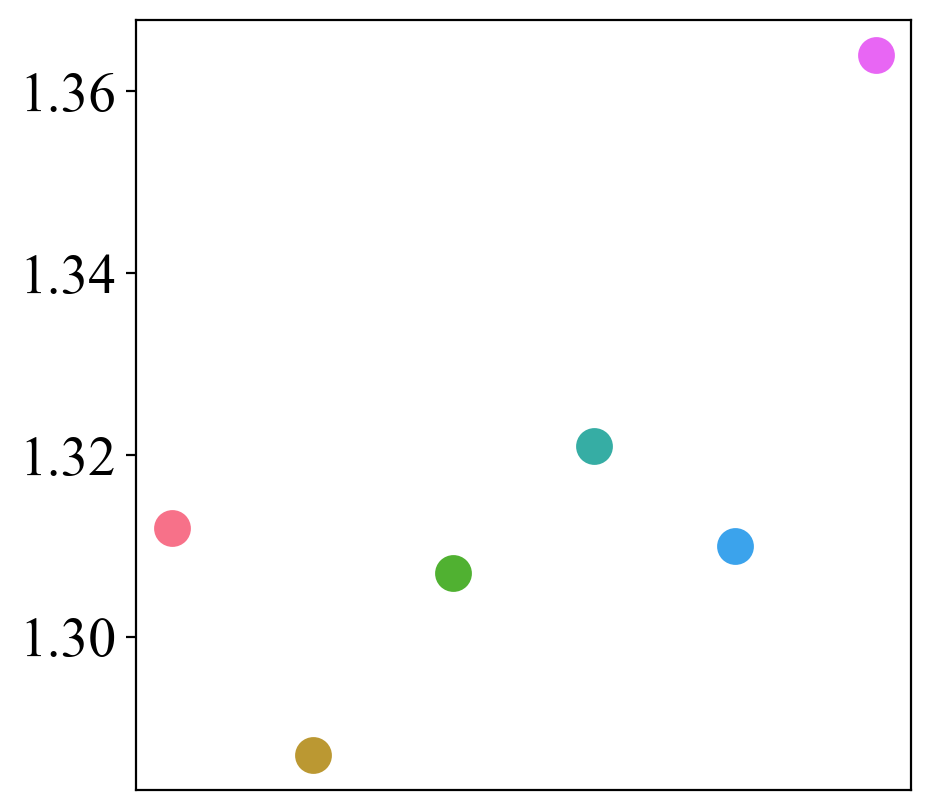}
         \caption{RMSE in daily mean temperature for extreme heat days}
         \label{fig:alt_rmse_q90}
     \end{subfigure}
     \begin{subfigure}[t]{0.15\textwidth}
         \includegraphics[width=\textwidth]{figures/rmse_legend.png}
         \vspace{0.4\textwidth}
     \end{subfigure}\\
     \hspace{-0.01\textwidth}
     \begin{subfigure}[t]{0.04\textwidth}
         \vspace{-7.5\textwidth}
         \includegraphics[width=\textwidth, angle=90]{figures/W_label.png}
     \end{subfigure}
     \hspace{-0.01\textwidth}
     \begin{subfigure}[b]{0.4\textwidth}
         \centering
         \includegraphics[width=\textwidth]{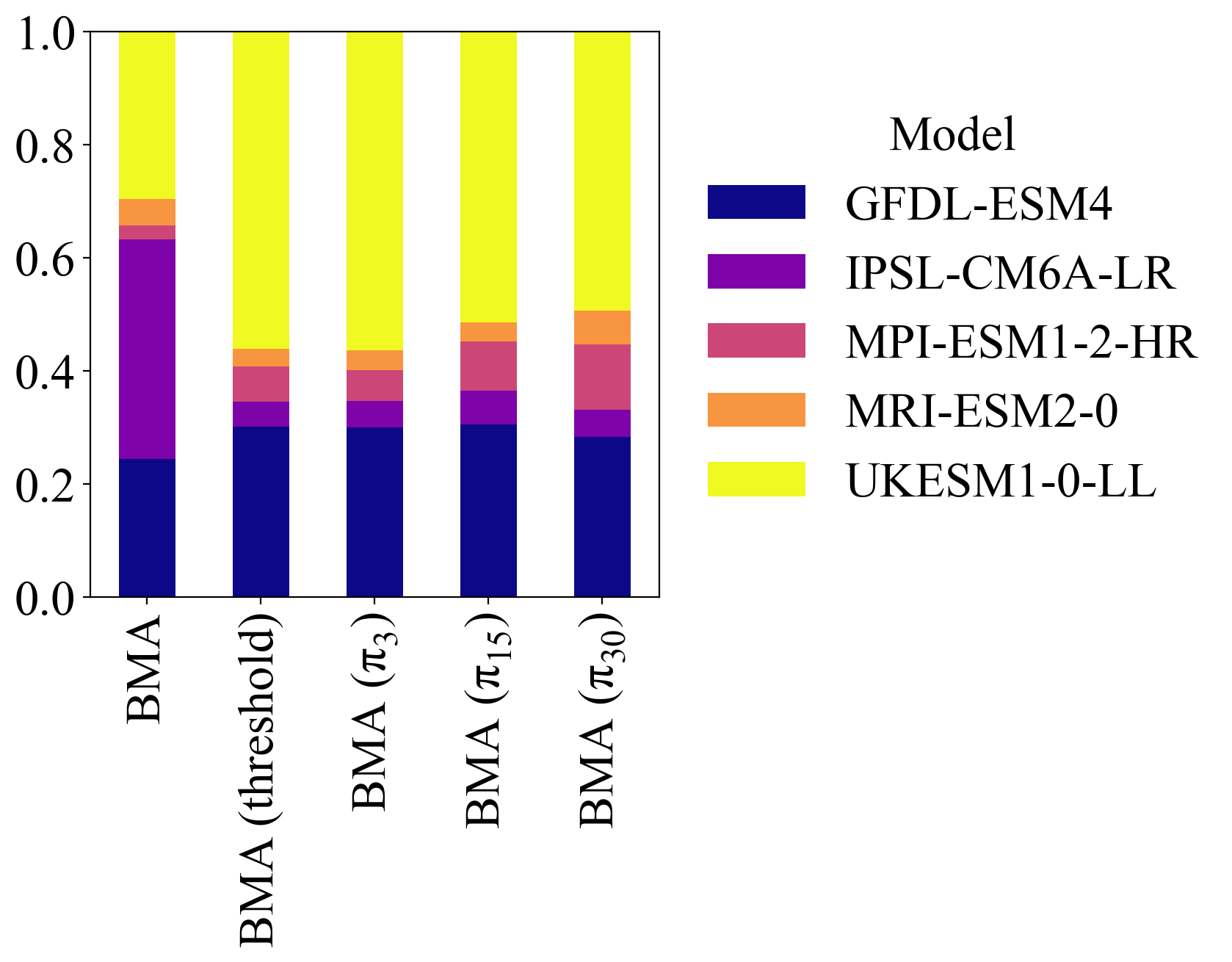}
         \caption{Climate model weights calculated from five BMA methods}
         \label{fig:alt_weights}
     \end{subfigure}
     \caption{Results for six ensemble methods for Nairobi using alternative model realisations, showing \ref{fig:alt_rmse_all}: RMSE for predicting daily average temperature for all days,  \ref{fig:alt_rmse_q90}: RMSE for predicting daily average temperature for extreme heat days, and \ref{fig:alt_weights}: model weights from BMA methods}
     \label{fig:alt_realisations}
\end{figure}

An additional set of experiments for Nairobi using an arbitrarily-chosen alternative realisation from each climate model was conducted. These realisations were GFDL-ESM4 \texttt{r1i1p1f1}, IPSL-CM6A-LR \texttt{r4i1p1f1}, MPI-ESM1-2-HR \texttt{r2i1p1f1}, MRI-ESM2-0 \texttt{r5i1p1f1} and UKESM1-0-LL \texttt{r2i1p1f2} (note that an alternative realisation for GFDL-ESM4 was unavailable). The results for RMSE for all days, RMSE for extreme heat days, and model weights assigned from each ensemble method from this experiment are shown in Figure \ref{fig:alt_realisations}.


\end{document}